\documentclass[aps,prb,twocolumn,groupedaddress,floatfix]{revtex4}
\usepackage{epsfig}
\usepackage{graphicx}
\usepackage{amsmath}

\begin{document}

\title{Short-circuit boundary conditions in ferroelectric PbTiO$_3$ thin films}

\author{Alexie M. Kolpak} 
\author{Na Sai} 
\author{Andrew M. Rappe}
\email{rappe@sas.upenn.edu} 
\affiliation{The Makineni Theoretical
Laboratories, Department of Chemistry, University of Pennsylvania,
\\231 S. 34th Street, Philadelphia, PA 19104--6323}

\date{\today}

\begin{abstract}
We examine the application of short-circuit electrical boundary conditions in density functional theory calculations of ferroelectric thin films. Modeling PbTiO$_3$ films with Pt electrodes, we demonstrate that under periodic boundary conditions of supercells, short-circuit conditions for the electrodes are equivalently satisfied in two repeated slab geometries: a PbTiO$_3$/metal superlattice geometry and a periodic metal/PbTiO$_3$/metal/vacuum geometry, where the metal is Pt or SrRuO$_3$. We discuss the benefits of each geometry in the study of ferroelectricity in thin films.
\end{abstract}
\maketitle

The application of relevant electrostatic boundary conditions in first-principles simulations of ferroelectric thin films is extremely important.~\cite{Meyer01p205426,Ghosez00p2767,Sai05p020101R,Junquera03p506,Almahmoud04p220102} In many studies, the calculations are carried out on systems composed of a ferroelectric thin film sandwiched between two metal electrodes in short-circuit conditions,~\cite{Shaw00p263,Dawber05p1083}i.e., the electrostatic potential through both electrodes is the same. The short-circuit boundary conditions are also relevant to most experiments.~\cite{Indlekofer05p282}

\begin{figure}
\includegraphics[angle=0,width=6cm]{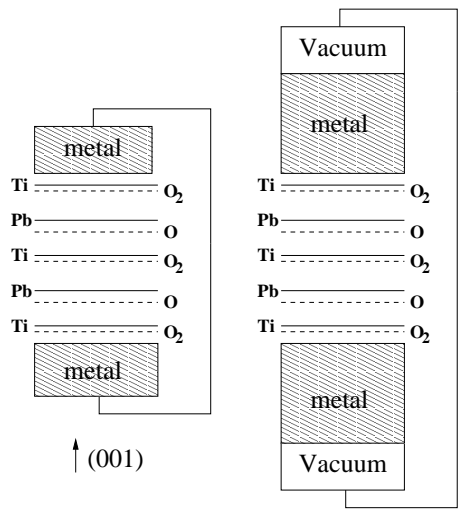}
\caption{Schematic of PbTiO$_3$ in the continuous (left) and isolated
(right) geometries.  Periodic slabs in the former
are in contact to form a continuous lattice, while in the latter the
slabs are separated by a large vacuum. The horizontal solid and
dashed lines indicate the relaxed positions of the cations and oxygens,
respectively, in the film surface normal direction.}
\label{fig:structures}
\end{figure}

Under supercell periodic boundary conditions in the first-principles studies, one way to satisfy the short-circuit conditions is to consider a slab geometry consisting of repeated ferroelectric/metal units, in which the metal layers are thick enough so that copies of the ferroelectric film do not interact with each other.~\cite{Junquera03p506} In this geometry, illustrated schematically in Fig.~\ref{fig:structures}a, the short-circuit boundary conditions are directly enforced, with contact between the metal electrodes at the top and bottom of the ferroelectric film through a bulk-like region of the metal.  In this case, the entire ferroelectric/electrode structure can be relaxed by considering the unit cell strain as a variational parameter.

It is also possible to model ferroelectric thin films using an isolated capacitor geometry that contains electrode/ferroelectric/electrode slabs that are separated by vacuum, such as in Fig.~\ref{fig:structures}b.  This geometry is computationally less efficient than the continuous geometry (requiring more atoms and larger supercells), but it is beneficial in that there is no artificial constraint on the unit cell strains along the surface normal direction, and therefore no need to rely on the strains to find the relaxed geometry. By separating the effects of the positive and negative ferroelectric surface charges on the metal behavior, an isolated geometry also facilitates the study of work functions at the ferroelectric surface~\cite{Sai05p020101R} and the comparison of complete and partial screening in thin metal films. Furthermore, this arrangement is useful for studying the effects of having different top and bottom electrodes, since it eliminates the influence of an additional inter-electrode interface. However, unlike in the continuous geometry, it is not obvious whether the necessary electrostatic boundary conditions can be achieved in an isolated structure.

In this paper, we demonstrate using density-functional
theory (DFT) calculations that short-circuit boundary conditions are
satisfied in both continuous and isolated geometries. We determine
the minimum electrode thickness in each geometry for ultrathin
ferroelectric PbTiO$_3$~(001) films with platinum (Pt) and SrRuO$_3$ electrodes, and
we demonstrate that, given sufficiently thick electrodes, the desired
electrostatic boundary conditions are satisfied in both geometries.
Furthermore, we show that with electrodes at or above the minimum
thickness, the ferroelectric structure of the perovskite films, as
well as the electronic behavior of both the metallic and perovskite
regions, is indistinguishable in the two geometries.

Our calculations were performed using DFT with the generalized
gradient approximation as implemented in the {\em ab initio} code
dacapo~\footnote{http://dcwww.camp.dtu.dk/campos/Dacapo/}, with a
plane wave cutoff of 30~Ry and a 4$\times$4$\times$1 Monkhorst-Pack
$k$-point mesh~\cite{Monkhorst76p5188}, and all calculations
correspond to $T$=0 K. The PbTiO$_3$ structures considered consist of
five atomic layers, three layers of TiO$_2$ separated by PbO planes
stacked in the (001) direction, as illustrated by the schematic in
Fig.~\ref{fig:structures}.  Experimentally, cubic SrTiO$_3$ and
tetragonal PbTiO$_3$ have the same in-plane lattice constant
(3.905~\AA), so the interface is stress free.  To model stress-free
growth of tetragonal PbTiO$_3$ on SrTiO$_3$, the in-plane lattice
constant was fixed to the theoretical equilibrium bulk value of
3.86~\AA.  All atomic coordinates were fully relaxed until the forces
on each atom are less than 0.01 eV/ A.  Polarization is perpendicular
to the interface, which is energetically favorable for this choice of
in-plane lattice constant.~\footnote{In fact, we find that
perpendicular polarization is favored over in-plane polarization for
PbTiO$_3$ in-plane lattice constants from 2.5\% less than the
equilibrium value up to 0.5\% above the equilibrium lattice constant,
at which point the in-plane polarization becomes more favorable.
Since the Pt(100) lattice constant is only 0.38\% larger than that of
SrTiO$_3$, we predict perpendicular polarization for PbTiO$_3$ grown
on Pt(100).  The in-plane polarization has been studied by I. Kornev
{\em et al.}  (Phys. Rev. Lett. {\bf 93}, 196104 (2004) and by Meyer,
Padilla, and Vanderbilt (Faraday Discussions {\bf 114}, 395
(1999)). We here address the out-of-plane polarization while leaving
the boundary conditions relating to in-plane polarization in a future
study.}  The interfacial Pt atoms are in the energetically-favored
positions above the TiO$_2$ oxygens, and additional Pt layers continue
the (001) lattice.  In the continuous geometry, periodic copies of the
PbTiO$_3$ film are separated by a Pt electrode of $l_{\rm Pt}=3$, 5,
7, or 9 atomic layers~\footnote{We consider electrodes with an odd
number of metal layers and a single middle layer, as required by the
continuous geometry.}, while in the isolated geometry, the PbTiO$_3$
film is sandwiched between two Pt electrodes, each of thickness
$l_{\rm Pt}$, and the entire structure is then separated from its
periodic images by 20~\AA\ of vacuum.  In the isolated geometry, a
dipole correction~\cite{Bengtsson99p12301} was applied in the center
of the vacuum region.

The ferroelectric behavior of the PbTiO$_3$ films can be characterized
by the set of rumpling parameters, $\delta z = z_{\rm Pb (Ti)}-z_{\rm
O}$, which describes the cation-oxygen separation along the
polarization direction for each PbO and TiO$_2$ layer in a given film.
In continuous structures, we expect that $\delta z$ will evolve as
$l_{\rm Pt}$ increases, until the Pt thickness is sufficient to
completely screen the two PbTiO$_3$ surfaces from each other.  We
expect the isolated geometry to behave similarly, with $\delta z$
changing until the electrodes are thick enough to prevent interactions
between each PbTiO$_3$ surface and the corresponding electrode/vacuum
interfaces.  Furthermore, we expect the converged $\delta z$ values to
be equal in the two geometries.

To determine the minimum electrode thickness for converged
ferroelectric thin film properties, we consider the evolution of
ferroelectric rumpling $\delta z$ as a function of electrode thickness
$l_{\rm Pt}$.  Table~\ref{tab:rumpling} gives the values for the
continuous and isolated geometries, respectively, as a fraction of the
bulk rumpling parameters.  As the table shows, the ferroelectric
behavior of the PbTiO$_3$ films with Pt electrodes is identical in the
continuous and isolated geometries for structures with $l_{\rm
Pt}\ge7$, with all atoms in the same positions to within $<$0.001~\AA.
Similarly, the PbTiO$_3$ rumpling profile in the two geometries
converge to a single structure with SrRuO$_3$ electrodes of nine or
more layers.  The table shows that PbTiO$_3$ films with Pt electrodes
have polarization enhanced by $\approx$15~\% relative to bulk at the
positive surface and decreased by slightly more at the negative
surface, while the central layers of the film show bulk-like rumpling.
The $\delta z$ values for films with SrRuO$_3$ electrodes indicate a
suppression of the polarization in all layers, with rumpling
parameters of $\approx$~20-35~\% of the bulk values.

\begin{figure*}
\includegraphics[angle=0,width=14cm]{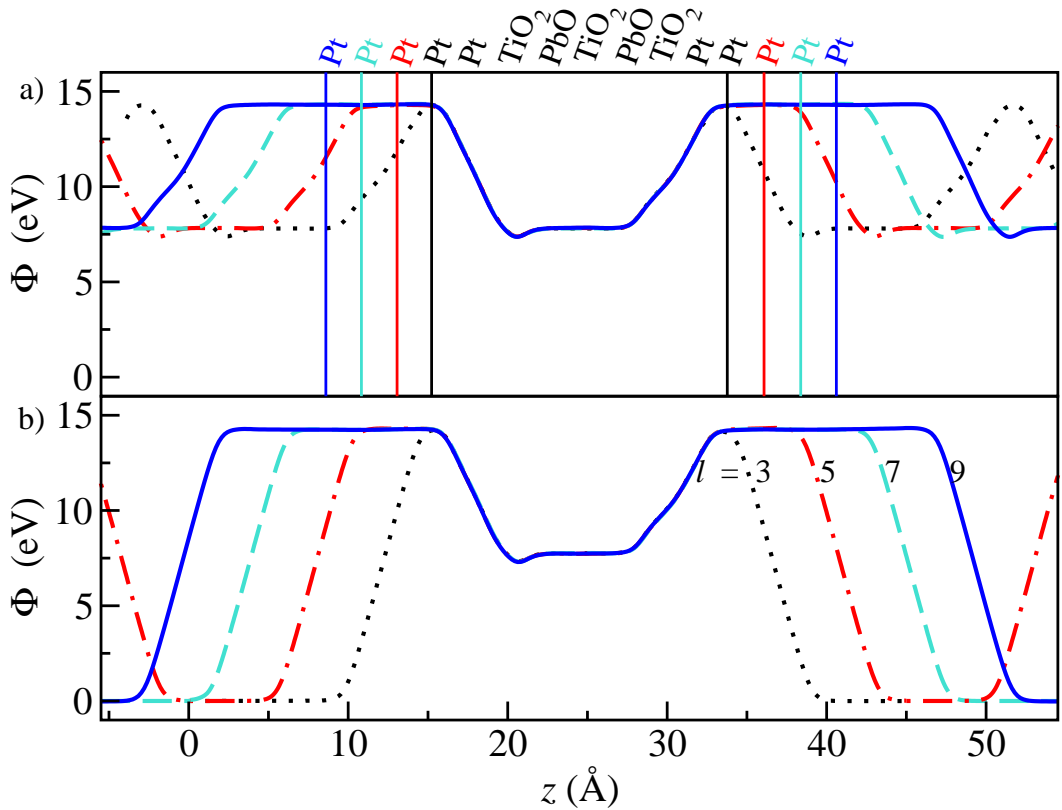}
\caption{Layer-averaged electrostatic potentials, $\Phi(z)$, along (001) for structures in the a) continuous and b) isolated geometries.  The curves for films with $l_{\rm Pt}=3$, 5, 7, and 9 are shown with black dotted, red dot-dashed, green dashed, and blue solid lines, respectively. The vertical black, red, green, and blue lines in a) bound a single repeat of the periodic supercell for $l_{\rm Pt}=3$, 5, 7, and 9, respectively; the lines on the left side of the ferroelectric film are positioned at the central Pt layer for each $l_{\rm Pt}$ (labeled in the corresponding color), and those on the right mark the same Pt layer in the next supercell.}
\label{fig:potentials}
\end{figure*}

In a grounded PbTiO$_3$/metal system, electrons are able to flow through
the electrode(s) from one ferroelectric/electrode interface to the
other, regardless of geometry.  Consequently, an electric field cannot
be sustained through the electrode, and therefore the electrostatic
potential must be flat through the electrode region.
Macroaveraging~\cite{Baldereschi88p734,Junquera03p506,Sai05p020101R}
the electrostatic potential in the continuous geometry, as illustrated
in Fig.~\ref{fig:potentials}a, shows that structures
with electrodes of seven and nine atomic layers of Pt are indeed in short-circuit,
with clear zero-slope regions through the center of the electrode.  On
the other hand, the structures with $l_{\rm Pt}<$~7 do not exhibit
rigorously equipotential regions.  As the potential for $l_{\rm Pt}=3$
illustrates, the Pt regions corresponding to each interface meet, leaving
no equipotential region.  Increasing the Pt thickness to five layers
allows separate Pt and interface regions to emerge, but a small
electrostatic potential slope remains, demonstrating that $l_{\rm
Pt}=5$ is insufficient to completely prevent interactions between the
periodic copies of the PbTiO$_3$ film.  The small slope through the
electrode for each $l_{\rm Pt}$ is given in the left side
of Table~\ref{tab:slopes}, quantitatively indicating the rapid
decrease of the field with increasing electrode thickness.  The zero
slope through the electrode for films with $l_{\rm Pt}>5$ confirms
that these structures are in short-circuit.

\begin{table}
\caption{Rumpling parameters $\delta z$ for the continuous and isolated geometries as a fraction of the theoretical bulk rumpling parameters,  $\delta z_{\rm bulk}$(TiO$_2$)=0.51 and $\delta z_{\rm bulk}$(PbO)=0.87 \AA.  The first TiO$_2$ layer corresponds to the bottom PbTiO$_3$ surface in Fig.~\ref{fig:structures}, with the surface normal parallel to the polarization. The ferroelectric-paraelectric energy difference, $\Delta E = E_{\rm ferro}-E_{\rm para}$, is also shown, in units of eV.  A negative value of $\Delta E$ indicates an energetically favorable ferroelectric state.}
\begin{center}
\begin{tabular*}{0.48\textwidth}{@{\extracolsep{\fill}}ccccccc}
\hline
$l_{\rm Pt}$ &TiO$_2$ &PbO  &TiO$_2$ &PbO  &TiO$_2$ &$\Delta E$\\\hline
{\em continuous}\\
3   &0.82  &0.88  &0.93  &0.93  &1.14 &-0.33 \\
5   &0.81  &0.91  &0.93  &0.94  &1.14 &-0.27 \\  
7   &0.82  &0.95  &0.96  &0.98  &1.16 &-0.24 \\  
9   &0.82  &0.95  &0.96  &0.98  &1.16 &-0.24 \\
{\em isolated}\\
3   &0.83  &0.95  &0.97  &0.99  &1.17 &-0.30 \\
5   &0.83  &0.94  &0.97  &0.99  &1.16 &-0.24 \\  
7   &0.82  &0.95  &0.96  &0.98  &1.16 &-0.24 \\  
9   &0.82  &0.95  &0.96  &0.98  &1.16 &-0.24 \\\hline
$l_{\rm SrRuO_3}$ &TiO$_2$ &PbO  &TiO$_2$ &PbO  &TiO$_2$ \\\hline
{\em continuous}\\
5   &0.38  &0.24  &0.36  &0.24  &0.30 &-0.09 \\  
7   &0.36  &0.22  &0.35  &0.22  &0.29 &-0.13 \\  
9   &0.35  &0.21  &0.34  &0.22  &0.28 &-0.14 \\
11  &0.36  &0.21  &0.33  &0.22  &0.28 &-0.14 \\
{\em isolated}\\
5   &0.34  &0.20  &0.32  &0.21  &0.26 &-0.01 \\  
7   &0.35  &0.21  &0.35  &0.22  &0.27 &-0.07 \\  
9   &0.36  &0.21  &0.33  &0.22  &0.27 &-0.14 \\
11  &0.36  &0.21  &0.33  &0.22  &0.28 &-0.14 \\
\hline
\end{tabular*}
\end{center}
\label{tab:rumpling}
\end{table}

The electrostatic potentials of PbTiO$_3$/Pt structures in the
isolated geometry, plotted in Fig.~\ref{fig:potentials}b, show a
similar picture.  As in the continuous geometry, it is not possible to
clearly differentiate the electrode from the interfaces for $l_{\rm
Pt}=3$.  In addition, five Pt layers are again not thick enough to
eliminate fields through the electrode due to the polarization charge,
as shown by the non-zero value of the slope given in
Table~\ref{tab:slopes}.  Most significantly, however, the potentials
are again flat through the central region of the electrodes for the
structures with $l_{\rm Pt}>5$, suggesting that charge is successfully
transferred between the two PbTiO$_3$/Pt interfaces.  The magnitude of
the dipole correction, $p_{\rm corr}$, also reported in
Table~\ref{tab:slopes}, further indicates that the system is in
short-circuit; $p_{\rm corr}$ goes to zero for $l_{\rm Pt}>5$, as
there is no net dipole through the Pt/PbTiO$_3$/Pt structure.  

Fig.~\ref{fig:potentials} also shows that the field through
the ferroelectric slab, $E_{\rm FE}$, is very similar for all $l_{\rm
Pt}$, regardless of the geometry.  The magnitude of the field through
the central unit cell (the middle three layers in
Fig.~\ref{fig:structures}), reported in Table~\ref{tab:slopes}, again
indicates that the minimum electrode thickness for converged
ferroelectric properties in both geometries is $l_{\rm Pt}=7$.
Interestingly, the evolution of $E_{\rm FE}$ with Pt thickness is
dependent on the geometry, with $E_{\rm FE}$ decreasing towards the
converged value in the continuous geometry, and increasing to the
converged value in the isolated geometry.  The difference in the trend
demonstrates the importance of choosing electrodes of sufficient
thickness in studying the behavior of charge passivation in
ferroelectric/metal capacitors.  The identical converged values of
$E_{\rm FE}$, however, show that the two geometries are equivalent
once this condition is satisfied.

With SrRuO$_3$ electrodes, the field through the
PbTiO$_3$ is significantly larger than that found with Pt electrodes.
Nevertheless, as Table~\ref{tab:slopes} shows,
the evolution of the electrostatic potential as a function of
electrode thickness is similar for both electrode materials.  In the
PbTiO$_3$/SrRuO$_3$ films, the slope through the ferroelectric
converges to a value of 0.053 eV/\AA\ and the field through the
electrodes approaches zero as the SrRuO$_3$ thickness is increased to
$l_{\rm SrRuO_3}=11$, regardless of the film geometry.  The larger
minimum electrode thickness determined for the PbTiO$_3$/SrRuO$_3$
films reflects the longer screening length of SrRuO$_3$ compared to
Pt.

The above data demonstrate that for both electrode materials, charge
transfer occurs between the two PbTiO$_3$/metal interfaces in the
isolated geometry, despite the fact that the electrodes are not in direct
contact.  This is possible due to the nature of the computation, which
requires that the electrodes are provided with electrons from a single
reservoir; i.e., there is a constant chemical potential of electrons.
Consequently, the minimum electrode thickness necessary to
short-circuit the system is defined in the same manner as in the
continuous geometry, as the thickness necessary to screen interactions
between interfaces, in this case, the PbTiO$_3$/metal and metal/vacuum
interfaces sandwiching each electrode.

\begin{table}
\caption{Magnitude of the electric fields $E_{\rm metal}$ and $E_{\rm
FE}$ through the electrode and ferroelectric regions, respectively, as
a function of electrode thickness for continuous and isolated
geometries.  The values for $E_{\rm metal}$ in the isolated geometry
show the average of the slope through the two electrodes.  The
magnitude of the dipole correction in the isolated geometry, $p_{\rm
corr}$, is also given, in units of e\AA. The electric fields, in units
of V/\AA, are determined from the macroaveraged\cite{Baldereschi88p734} electrostatic potentials by a linear fit to the central 4~\AA\ in the
PbTiO$_3$ films and the SrRuO$_3$ electrodes, and to the inner $l$-2
layers in the Pt electrodes.  The uncertainty in the electric fields is $\pm 0.0005$ V/\AA.}
\begin{center}
\begin{tabular*}{0.48\textwidth}{@{\extracolsep{\fill}}|c|c|c|c|c|c|}
\hline
&\multicolumn{2}{c|}{continuous} &\multicolumn{3}{c|}{isolated}\\ \hline
$l_{\rm Pt}$ &$E_{\rm Pt}$ &$E_{\rm FE}$ &$E_{\rm Pt}$ &$E_{\rm FE}$ &$p_{\rm corr}$ \\
\hline
3   &         &0.0100   &         &0.0031   &0.007   \\
5   &0.0103   &0.0006   &0.0180   &0.0006   &0.003   \\
7   &0.0001   &0.0004   &0.0001   &0.0003   &0.001   \\
9   &0.0001   &0.0003   &0.0001   &0.0003   &0.000   \\\hline
$l_{\rm SrRuO_3}$ &$E_{\rm SrRuO_3}$ &$E_{\rm FE}$ &$E_{\rm SrRuO_3}$ &$E_{\rm FE}$ &$p_{\rm corr}$ \\\hline
5   &0.0350  &0.0624  &0.0838  &0.0460  &0.026  \\
7   &0.0029  &0.0549  &0.0160  &0.0520  &0.015  \\
9   &0.0001  &0.0525  &0.0002  &0.0534  &0.000  \\
11  &0.0001  &0.0530  &0.0001  &0.0530  &0.000  \\
\hline
\end{tabular*}
\end{center}
\label{tab:slopes}
\end{table}

\begin{figure}
\includegraphics[angle=0,width=6cm]{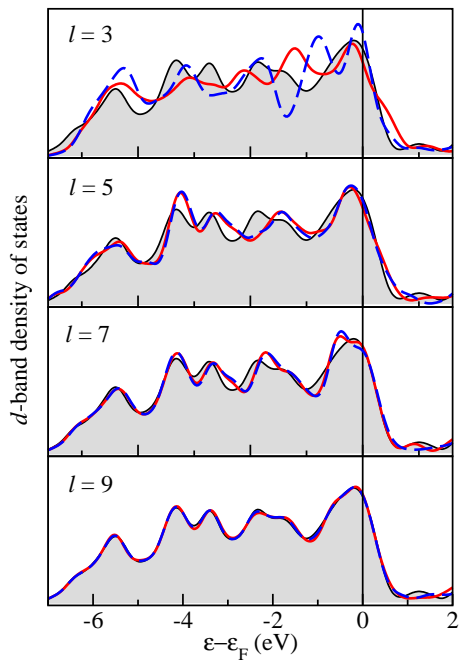}
\caption{$d$-band DOS of the central Pt layer in the electrodes of the
continuous (solid red curve) and isolated (dashed blue curve)
geometries.  In both geometries, the Pt DOS in the center of the
nine-layer electrodes is virtually identical to that in the center of a
thick Pt~(001) slab (shaded).  }
\label{fig:DOS}
\end{figure}

Examination of the density of states (DOS) of the central metal layer
offers further confirmation that short-circuit boundary conditions
have been satisfied in both geometries.  In grounded systems with
electrodes of sufficient thickness to prevent interaction between
periodic copies of the PbTiO$_3$ film, the electronic structure of the
central metal layer should be identical to the corresponding
layer in an (001) metal slab, as the effects of the
ferroelectric film (or the vacuum) will be completely screened in this
region.  In Fig.~\ref{fig:DOS}, the $d$-band DOS of PbTiO$_3$/Pt
structures in the continuous and isolated geometries are compared to
the DOS of the middle Pt layer in a 9-layer Pt~(001) film with the
same in-plane lattice constant.  As the top panel of the figure shows,
the $d$-band of the continuous (solid curve) and isolated (dashed
curve) structures with $l_{\rm Pt}=3$ differ significantly from the
Pt~(001) $d$-band (shaded).  They also differ from each other, as
expected from the structural data shown in Table~\ref{tab:rumpling}.
As $l_{\rm Pt}$ is increased, the continuous and isolated $d$-bands
look increasingly similar to that of the Pt~(001).  Although there are
still minor differences between the DOS of the 7-layer structures and
Pt~(001), the isolated and continuous $d$-bands for the $l_{\rm Pt}=9$
structures are both essentially indistinguishable from the Pt~(001)
$d$-band.

The DOS of the central layer in the SrRuO$_3$ electrodes, not shown
due to space constraints, also becomes very similar to the
corresponding SrO or RuO$_2$ layer in a thick SrRuO$_3$ slab as
$l_{\rm SrRuO_3}$ increases, displaying only small differences for 
$l_{\rm SrRuO_3}\ge9$.

In conclusion, we have shown that short-circuit boundary conditions
are satisfied in ferroelectric PbTiO$_3$/metal capacitor structures in
two geometries when the electrode thickness is greater than five Pt
layers or seven SrRuO$_3$ layers.  We have shown that, for a given electrode material, the ferroelectric
films in the continuous and isolated geometries converge to the same
atomic structure, and that the
potential in the ferroelectric and metal regions is identical for both
geometries with electrodes of sufficient thickness, resulting in
indistinguishable ferroelectric behavior.  In such structures, the
electrostatic potential has zero slope through the center of the metal
electrode, demonstrating that electron transfer between the
PbTiO$_3$/metal interfaces is accomplished, via direct contact between
the electrodes in the continuous geometry or as a result of a single
reservoir of electrons shared by the electrodes in the isolated
geometry.  An electronic structure analysis shows that the DOS of the
central metal layer is almost identical to that in the corresponding
metal slab, further demonstrating that these structures are in
short-circuit.  Our results demonstrate that ferroelectric capacitor
structures can be accurately modeled in either geometry, allowing the
possibility to utilize the benefits of both in further studies.

This work was supported by the Office of Naval Research under grant
number N-000014-00-1-0372, the Center for Piezoelectric Design, and
the National Science Foundation, through the MRSEC program, grant
No. DMR00-79909.  Computational support was provided by the HPCMO and
DURIP.  AMK is supported by an Arkema Inc.\ fellowship.

\bibliography{other-rappecites}
\end{document}